\documentclass[aps,pre,preprint,groupedaddress]{revtex4}

\begin{document}


\title{Comment on an information theoretic approach to the study of
non-equilibrium steady states}

\author{Glenn C.~Paquette}
\email[]{paquette@scphys.kyoto-u.ac.jp}
\affiliation{Department of Physics, Graduate School of Science, Kyoto University, Kyoto 606-8502, Japan}

\date{\today}

\begin{abstract}
We argue that there is a fundamental problem regarding the analysis that serves as the foundation
for the papers
{\it Information theory explanation of the fluctuation 
theorem, maximum entropy production and 
self-organized criticality in non-equilibrium stationary 
states} [R.~Dewar, J.~Phys.~A: Math.~Gen.~{\bf 36} (2003), 631--641] and {\it Maximum entropy production
and the fluctuation theorem}
[R.~Dewar, J.~Phys.~A: Math.~Gen.~{\bf 38} (2005), L371--L381]. In particular, we demonstrate that this
analysis is based on an assumption that is physically unrealistic and that, hence, the results
obtained in those papers cannot be regarded as physically meaningful.
\end{abstract}


\maketitle

In this paper, we consider two works
\cite{dewar1, dewar2} that have been quite influential in recent studies in the fields of atmospheric science, environmental science and ecology,
particularly in connection to the so-called principle of maximum entropy production \cite{kleidon}. We find that these works are seriously flawed.
Specifically, we argue that the variational derivation on which they are based begins with the assumption of a condition that is physically unfeasible
and that, thus, although the computation itself is correct, its result lacks physical meaning. 
We then provide a particular example that demonstrates this point explicitly.

In Ref.~\cite{dewar1}, the author derives an expression (Eq.~(5) there) that is claimed to be the probability distribution for the microscopic trajectories
of a general open system. This probability distribution is the fundamental result on which
Refs.~\cite{dewar1} and \cite{dewar2} are based. He then proceeds to derive from this fundamental result a number of secondary results (the fluctuation theorem,
a condition of maximum entropy production as the selection principle for non-equilibrium steady states, behavior representing the emergence of self-organized
criticality, and relations that indicate the connection between the fluctuation theorem and the maximum entropy production selection principle).
Several years after the appearance of
Refs.~\cite{dewar1} and \cite{dewar2}, there appeared two works that criticize the derivations of some of these secondary results.
Bruers \cite{bruers} pointed out a possible
problem involving an assumption made in Ref.~\cite{dewar1} that is needed to obtain the maximum entropy production selection principle. There it is claimed
that the proper assumption in fact leads to a prediction of minimum entropy production. 
Later, Grinstein and Linsker \cite{GL} elucidated two separate problems involving the use of
two approximations beyond their regimes of validity, in the derivation of the result representing self-organized criticality in Ref.~\cite{dewar1}, and
in the derivation of the relations connecting the fluctuation theorem to the non-equilibrium steady state selection principle in Ref.~\cite{dewar2}.
These are important points. However, they all regard analysis that appears subsequent to
the derivation of the fundamental result, Eq.~(5) of Ref.~\cite{dewar1}. These papers point out possible problems in the application of this
fundamental result, but they in no way call into question this result itself. By contrast, the problem demonstrated presently, which
regards the derivation of the fundamental result, is more serious and casts doubt on the merit of the information theoretic approach proposed in Ref.~\cite{dewar1}.

As stated above, in Ref.~\cite{dewar1}, the author attempts to derive the stationary probability distribution for the (classical) microscopic
trajectories of a general open system, exchanging energy and particles with its environment.
He does this using a variational approach, assuming that it can be accomplished by maximizing
the generalized entropy $-\sum_i P(\Gamma_i) \log P(\Gamma_i)$,
subject to the proper conditions. (Note that to suit the present purposes, here and below we use notation
that differs slightly from that in the original.)
Here, $\Gamma_i$ represents the $i$th trajectory between some specified initial and
final times (which we choose as $t =0$ and $t = \tau$),
$P(\Gamma_i)$ is its probability, and the sum is over all possible trajectories between these times. 
The author assumes that in order to obtain the correct distribution, the only physical quantities to which constraints must be applied are the
energy density and particle densities at each point in the system, expressed collectively as $d(x,t)$, and the energy flux and particle fluxes
at every point on the boundary of the system, expressed as $F(x,t)$. In fact, as the constraints imposed in the variation,
the author uses only the following:
\begin{eqnarray}
\sum_i P(\Gamma_i) &=& 1 \; , \nonumber \\
\sum_i P(\Gamma_i) d_i(x,0) &=& A(x) \; {\rm{for}} \; x \in V \; , \nonumber \\
\sum_i P(\Gamma_i) \overline{F}_i(x) &=& B(x) \; {\rm{for}} \; x \in \Omega \; .
\end{eqnarray}
Here, $d_i(x,0)$ is the initial value of $d(x,t)$ for the $i$th trajectory,
$\overline{F}_i(x)$ is the time average of $F(x,t)$ taken over the entire $i$th trajectory, $A(x)$ and $B(x)$ are
some specified functions, fixing the ensemble averages $\langle d_i(x,0) \rangle$ and $\langle \overline{F}_i(x) \rangle$,
and $V$ and $\Omega$ represent the system and its boundary, respectively.
In Refs.~\cite{dewar1} and \cite{dewar2}, it is claimed that the proper form of $P(\Gamma_i)$ can be obtained by maximizing the above generalized entropy subject to these constraints
alone.

The most important point to note here is that in the proposed variation, nothing distinguishes between
two arbitrary trajectories $\Gamma_i$ and $\Gamma_j$
that satisfy the relations $d_i(x,0) = d_j(x,0)$ and $\overline{F}_i(x) = \overline{F}_j(x)$.
Hence, it is implicitly assumed that the probabilities of any two trajectories with
identical initial conditions and average fluxes are equal. This is seen clearly in
the resulting probability distribution, appearing in Eq.~(5) of Ref.~\cite{dewar1},
$P(\Gamma_i) = Z^{-1}\exp[-A(\Gamma_i)]$ with $Z \equiv \sum_i \exp[-A(\Gamma_i)]$ and
$A(\Gamma_i) = \int_V \lambda(x) \cdot d_i(x,0) dx + \int_\Omega \eta(x) \cdot \overline{F}_i(x) dx$, where $\lambda(x)$ and $\eta(x)$
are Lagrange multipliers. Although this is not the final form of $P(\Gamma_i)$ used there, subsequent manipulations serve only to
express the same quantity in a different manner; that is, there is no additional quantity introduced into $A(\Gamma_i)$
that would alter the assumption stated above.
On the basis of physical considerations, however, this appears to be an unrealistically strong assumption, as it
implies that a trajectory along which the energy or quantity of material possessed by a system fluctuates wildly and one along which it fluctuates very gently
will have equal probabilities as long as they have the same initial and final states. Below, we investigate this point by considering a particular system,
described by the prototypical model of thermal fluctuations. From this investigation we find that, indeed, the assumption made in Ref.~\cite{dewar1} is physically invalid.

In this paper, we investigate the validity of the approach presented in Ref.~\cite{dewar1} by considering the dynamics described by the following Langevin equation:
\begin{equation}
\dot{\alpha}(t) = - \zeta \alpha(t) + \xi(t) \label{langevin} \; .
\end{equation}
Here, $\alpha$ can be regarded as an arbitrary extensive quantity representing the state of a thermodynamic system that is contact with a (heat and/or particle)
reservoir. The interaction of the system with the reservoir is characterized by $\zeta$, a positive constant, and $\xi(t)$, a stochastic force whose statistics are independent of $\alpha$ and $t$. 
The Langevin equation considered here has been investigated for more than a century \cite{langevin}, and while it was originally studied as a model of
Brownian motion, its physical validity has been firmly established in many contexts.
It is regarded as a prototype of the stochastic models used to describe open systems, as  it represents perhaps the simplest model of fluctuations that result from
the accumulation of many microscopic processes. Due to its minimal nature within such a class of models, the behavior it describes is observed quite universally among
systems exhibiting fluctuations of this kind. Indeed, (\ref{langevin}) represents the generic equation of motion for a single
fluctuating thermodynamic quantity in the linear response regime  \cite{onsager}. (Although this assumes that spatial gradients
of $\alpha$ can be ignored, even if
they cannot be ignored, the essential nature of the description is unchanged \footnote{In the case that spatial gradients cannot be ignored,
we simply obtain the multi-mode generalization of (\ref{langevin}), with the components of the vector-valued $\alpha$
representing the values of this quantity in small spatial cells. Thus, while the details of the analysis presented below would become somewhat more complicated in that case, the important points and
the main conclusion would be unchanged.}.)
For example, the dynamics of $\alpha$ described by this equation can be used to model fluctuations undergone by the concentrations of the chemical species participating in
a chemical reaction within some region.

Here we consider the simple case in which $\xi$  is a zero-mean, Gaussian-distributed stochastic force with delta-function time correlation. We also assume that the
coefficient $\zeta$ and the intensity of $\xi$ are related by the fluctuation-dissipation theorem of the second kind. This implies that, for example, in the case of a Brownian particle,
we have $\langle \xi(t) \xi(t') \rangle = 2 \zeta T m \delta(t - t')$ (where $m$ is the mass of the Brownian particle and $T$ is the temperature of the heat bath), while in the case of a chemical reaction,
we have $\langle \xi(t) \xi(t') \rangle = 2 \zeta T  \left( \sum_{ij} \nu_i \nu_j \frac{\partial \mu_i}{\partial n_j}\right)^{-1}\delta(t-t')$ (where $\nu_i$, $\mu_i$ and $n_i$ are the stoichiometric coefficient, the chemical potential and the concentration of species $i$, respectively).

From this point, for convenience, we regard (\ref{langevin}) as describing a Brownian particle. However, it should be kept in mind that this analysis and the conclusion
to which it leads apply to a very wide range of systems, including spatially extended systems of  many kinds.

Let us first rewrite (\ref{langevin}) to make its present application clearer:
\begin{equation}
\dot{p} = - \frac{\gamma}{m} p + \xi \; . \label{langevin2}
\end{equation}
Here, $p$ is the momentum of the Brownian particle, $m$ is its mass, and $\gamma$ is the friction constant.
Because the system here consists only of the Brownian particle, which is treated as a point mass,
the quantities $d_i(x,0)$ and $\overline{F}_i(x)$ are simply the initial kinetic energy of the particle and its average rate of change over a trajectory.

We now derive the probability for a trajectory of this Brownian particle. We begin by noting that
in the general case of Gaussian noise,
the probability for
the realization of a particular value of $\xi$ is given by
\begin{equation}
P(\xi) = c\exp[- (\xi - \langle \xi \rangle)^2/2\langle ( \xi - \langle \xi \rangle )^2 \rangle] \; ,
\end{equation}
where $c$ is a normalization constant.
For a system of the type that we consider, this becomes
\begin{equation}
P(\xi) = c \exp[- dt\xi^2/4\gamma T] \; ,
\end{equation}
where $dt$ can be regarded as the timescale of the description. It is important here that $c$ depends only
on $dt$, $\gamma$ and $T$ (and, of course, the number of spatial dimensions). From this, noting the Markovian nature of the
system, we immediately obtain the
probability for a finite-length trajectory $\Gamma$, given the initial state $p(0)$, realized under a particular noise history,
\begin{equation}
P(\Gamma |p(0)) = C \exp[- \int_0^\tau dt\xi^2(t)/4\gamma T] \; ,
\end{equation}
where $C$ depends on $\tau$, $\gamma$ and $T$ only.

Now, let us consider the set of all trajectories satisfying $p(0) = p(\tau) =0$.
Clearly, for any $a \ge 0$, there exists a realization of $\xi(t)$ (in fact, infinitely many) such that $\int_0^\tau dt \xi^2(t) = a$
and the resulting trajectory is an element of this set, given the initial condition.
However, the total change in the energy of the system over each such trajectory is zero. Thus, for all of these trajectories, both the initial conditions and
the average fluxes are identical. Nevertheless, two such trajectories with distinct values of $\int_0^\tau dt \xi^2(t)$
possess different probabilities. This contradicts the assumption on which  the derivation of $P(\Gamma)$ in Ref.~\cite{dewar1} is based.

To explicitly demonstrate this point, let us consider the set of trajectories for which $p(t)$ takes the form
$p(t) = A \sin(\pi t/\tau)$.
Then, substituting $\dot{p} + \frac{\gamma}{m}p$ for $\xi$ in the above integral, a trivial calculation yields 
\begin{equation}
\int_0^\tau dt \xi^2 = \left[ \frac{\pi^2}{\tau} + \left(\frac{\gamma}{m} \right)^2 \tau \right] A^2 \; .
\end{equation}
Thus, for distinct values of $|A|$, the corresponding trajectories have distinct values of $\int_0^\tau dt \xi^2$, and indeed
for any non-negative value of $\int_0^\tau dt \xi^2$, there exists a trajectory of the above form for which this value is realized.

The case considered above is that in which the asymptotic state of the system is equilibrium. Next, we note that the situation
is essentially unchanged if instead of (\ref{langevin2}) we consider the equation
\begin{equation}
\dot{p} = - \frac{\gamma}{m} p + f + \xi \; ,
\end{equation}
where $f$ represents a constant driving force, independent of space and time. For a system described by this equation with nonzero $f$, the asymptotic state is
a non-equilibrium steady state. All of the above analysis can be applied to this equation as well, and in this case, using the same form of $p(t)$ as above, we obtain
\begin{equation}
\int_0^\tau dt \xi^2 = \left[ \frac{\pi^2}{\tau} + \left(\frac{\gamma}{m} \right)^2 \tau \right] A^2 - \frac{4}{\pi} \frac{\gamma}{m} f \tau A + f^2 \tau \; .
\end{equation}
While we could consider further generalized forms of (\ref{langevin2}), it is sufficiently clear that the main conclusion in each case would be the same.


The problem of deriving probability distributions for non-equilibrium steady states has been studied for many years.
This is an extremely difficult problem, and to this time, except in certain simple cases, exact expressions -- even for single-time distribution functions -- have been obtained
only in non-explicit forms \cite{previous}. Noting that non-equilibrium steady states are realized as asymptotic fixed-point solutions,
it is natural to conjecture that such states, 
like equilibrium states, correspond to minima of some quantity characterizing the state space, 
i.e., some kind of non-equilibrium thermodynamic potential. Thus, the variational approach employed in Ref.~\cite{dewar1} is intuitively
appealing. However, the main difficulty involved in such an approach is that there exists no general principle that allows us to identify the
proper constraints to be used in the variational procedure. In the equilibrium case, the principle of equal
probability of micro states provides the information needed to obtain the single-time probability distribution. In the linear-response regime,
this principle, through the linear phenomenological laws, provides the same information for distributions of finite-length trajectories \cite{onsager}. However, because
we do not know of any such principles that apply outside of these limited regimes, the problem of obtaining generally valid probability distributions for
non-equilibrium steady states remains unsolved.

\begin{acknowledgments}
I am grateful to J.B.~Marston for useful discussions. This work was supported by the Grant-in-Aid for the Global COE Program
``The Next Generation of Physics, Spun from Universality and Emergence" from the Ministry of Education, Culture, Sports, Science and Technology (MEXT) of Japan.
\end{acknowledgments}

\end{document}